%% file: main.tex
\documentclass[conference]{IEEEtran}
\usepackage[T1]{fontenc}
\usepackage[latin9]{inputenc}

\usepackage{amsmath}
\usepackage{caption}
\usepackage{xpatch}
\usepackage{amssymb}
\usepackage{esint}
\usepackage{mathtools}
\usepackage{amsthm}
\usepackage{nicefrac}
\usepackage{bigints}
\usepackage{mathtools}
\usepackage{blkarray, bigstrut}
\usepackage{physics}
\usepackage{calligra}
\usepackage{graphicx}
\usepackage{epstopdf}
\usepackage{dsfont}
\usepackage{array,ragged2e}
\usepackage[acronyms,nonumberlist,nopostdot,nomain,nogroupskip]{glossaries}
\usepackage{enumitem}
\usepackage{etoolbox}
\usepackage{babel}
\usepackage[nopar]{lipsum}
\usepackage{psfrag}
\usepackage[ruled,lined,linesnumbered]{algorithm2e}
\usepackage{algorithmic}
\usepackage{algorithm2e}
\usepackage{balance}
\usepackage{float}
\usepackage{hyperref}

\SetKw{KwBy}{by}

\makeatletter
\newcommand{\removelatexerror}{\let\@latex@error\@gobble}
\makeatother

\graphicspath{ {Figures/} }

\xpatchcmd{\proof}{\hskip\labelsep}{\hskip5\labelsep}{}{}  
\makeatletter
\xpatchcmd{\proof}{\@addpunct{.}}{\@addpunct{:}}{}{}
\makeatother

\renewcommand\[{\begin{equation}}
\renewcommand\]{\end{equation}} 
\usepackage{listings}
\usepackage{fancyvrb}
\usepackage{framed}

\usepackage{courier}
\usepackage[dvipsnames,table]{xcolor}

\definecolor{dkgreen}{rgb}{0,0.3,0}
\definecolor{gray}{rgb}{0.5,0.5,0.5}

\makeatletter
\newcommand*{\rom}[1]{\expandafter\@slowromancap\romannumeral #1@}
\makeatother

\input{acronyms}

\usepackage{tabu}
\usepackage{booktabs}
\usepackage{soul,color}
\usepackage{multirow}
\usepackage{capt-of}
\usepackage{array}
\usepackage{arydshln}
\usepackage{cite}

\newcommand{\comment}[1]{}

\def\BibTeX{{\rm B\kern-.05em{\sc i\kern-.025em b}\kern-.08em
    T\kern-.1667em\lower.7ex\hbox{E}\kern-.125emX}}

\begin{document}
\addtolength{\topmargin}{0.03in}
\title{




Toward Generative 6G Simulation: An Experimental Multi-Agent LLM and ns-3 Integration

}

\author{\IEEEauthorblockN{
Farhad Rezazadeh\IEEEauthorrefmark{1},~Amir~Ashtari~Gargari\IEEEauthorrefmark{1},~Sandra~Lag\'en\IEEEauthorrefmark{1},~Houbing Song\IEEEauthorrefmark{2},~Dusit~Niyato\IEEEauthorrefmark{3},~and Lingjia Liu\IEEEauthorrefmark{4}
}

\IEEEauthorrefmark{1}\normalsize{}Centre Tecnol\'ogic de Telecomunicacions de Catalunya (CTTC), Barcelona, Spain\\

\IEEEauthorrefmark{2}University of Maryland, Baltimore County (UMBC), Baltimore, USA\\

\IEEEauthorrefmark{3}Nanyang Technological University, Singapore\\

\IEEEauthorrefmark{4}Virginia Tech, Blacksburg, USA\\

{\normalsize{}Contact Email:  
\texttt{farhad.rh@ieee.org}

}
}

\maketitle

\begin{abstract}
The move toward open Sixth-Generation (6G) networks necessitates a novel approach to full-stack simulation environments for evaluating complex technology developments before prototyping and real-world implementation. This paper introduces an innovative approach\footnote{A lightweight, mock version of the code is available on GitHub at \url{https://github.com/frezazadeh/LangChain-RAG-Technology}} that combines a multi-agent framework with the Network Simulator 3 (ns-3) to automate and optimize the generation, debugging, execution, and analysis of complex Fifth-Generation (5G) network scenarios. Our framework orchestrates a suite of specialized agents---namely, the Simulation Generation Agent, Test Designer Agent, Test Executor Agent, and Result Interpretation Agent---using advanced LangChain coordination. The Simulation Generation Agent employs a structured chain-of-thought (CoT) reasoning process, leveraging large language models (LLMs) and retrieval-augmented generation (RAG) to translate natural language simulation specifications into precise ns-3 scripts. Concurrently, the Test Designer Agent generates comprehensive automated test suites by integrating knowledge retrieval techniques with dynamic test case synthesis. The Test Executor Agent dynamically deploys and runs simulations, managing dependencies and parsing detailed performance metrics. At the same time, the Result Interpretation Agent utilizes LLM-driven analysis to extract actionable insights from the simulation outputs. By integrating external resources such as library documentation and ns-3 testing frameworks, our experimental approach can enhance simulation accuracy and adaptability, reducing reliance on extensive programming expertise. A detailed case study using the ns-3 5G-LENA module validates the effectiveness of the proposed approach. The code generation process converges in an average of 1.8 iterations, has a syntax error rate of 17.0\%, a mean response time of 7.3 seconds, and receives a human evaluation score of 7.5.

\end{abstract}


\section{Introduction}

\IEEEPARstart{T}{he} surge in demand for applications such as extended reality (XR) and holographic services is driving the rapid evolution toward 6G networks, which are set to integrate advanced technologies like \gls{thz} communication, AI-driven network management, and \glspl{ntn}~\cite{Ref5ns3,10680436}. These innovations require open interfaces and protocols to ensure seamless interoperability, yet the heterogeneous and complex nature of 6G demands sophisticated testing methods---particularly given the scarcity of full-scale 6G testbeds~\cite{Ref3ns3,10228969}.

A comprehensive, full-stack simulator is essential for validating novel methodologies across all network layers by using large-scale network deployments, before actual standardization, prototyping, and implementation. Tools such as ns-3 have historically enabled the evaluation of complex network environments~\cite{Ref1ns3,zubow2023toward}. The ns3 is a powerful and widely utilized network simulator that plays a critical role in developing and evaluating networks by modeling and assessing various approaches proposed by scholars~\cite{Ref2ns3}. However, the simulator's frequent updates and the inherent complexity of its C++ codebase can impede rapid prototyping and widespread adoption.

Recent advancements in language models have opened the door to automating the generation, debugging, and execution of simulation code from natural language descriptions~\cite{llamacoderef}. However, these models face challenges with complex tasks due to their limited contextual integration and dependency on pre-trained data. Hybrid methodologies such as retrieval-augmented generation (RAG) have emerged to address these limitations, enhancing generative models with external, domain-specific knowledge~\cite{ragrefren}. Using LLMs in the simulation process can help reduce the need for manual coding. This change can speed up innovation and make it easier for researchers working in dynamic 6G network environments.

\subsection{Contributions}

This paper presents a novel case study that integrates Multi-Agent LLMs with the ns-3 simulator to create a generative simulation framework explicitly tailored for 5G/6G networks. Our contributions are summarized as follows:

\begin{enumerate}
    \item \textbf{Multi-Agent Architecture:} We introduce a highly coordinated multi-agent system where specialized agents collaboratively manage the simulation lifecycle. The Simulation Generation Agent (Agent\#1) transforms natural language simulation requirements into executable code using state-of-the-art LLMs (via the ChatOpenAI interface) and prompt templates. In parallel, the Test Designer Agent (Agent\#2) constructs targeted test cases by leveraging retrieval-augmented techniques with a Pinecone vector store and OpenAIEmbeddings to ensure simulation accuracy.
    
    \item \textbf{Seamless Integration of LLMs with Simulation Tools:} Our framework bridges advanced LLM capabilities with domain-specific simulation tools. It dynamically orchestrates various APIs and tools such as the \texttt{CppSubprocessTool} for ns-3 C++ execution and the \texttt{PythonREPLTool} for Python debugging-allowing for automated code generation, execution, and iterative refinement.
    
    \item \textbf{Dynamic API and Model Optimization:} We implement a flexible toggle mechanism within the Streamlit interface that enables dynamic selection between optimized model variants (e.g., \texttt{gpt-4o-mini}) for cost-efficient operations. This integration ensures that our system maintains acceptable accuracy.
    
    \item \textbf{Case Study on 5G Network Simulation:} We validate our approach with a detailed case study focusing on 5G network scenarios. The case study demonstrates how the interplay between the Simulation Generation, Test Designer, Test Executor (Agent\#3), and Result Interpretation Agents (Agent\#4) leads to faster simulation development, robust debugging, and refined simulation outputs through iterative feedback loops.
\end{enumerate}

Overall, our work lays a robust foundation for integrating LLM-driven multi-agent systems with traditional simulation platforms, advancing the state-of-the-art automated network simulation and providing valuable insights for next-generation 6G network design and analysis.

\begin{figure}[t]
\centering
\includegraphics[width=.9\columnwidth, clip,trim={0cm 0cm 0cm 0cm}]{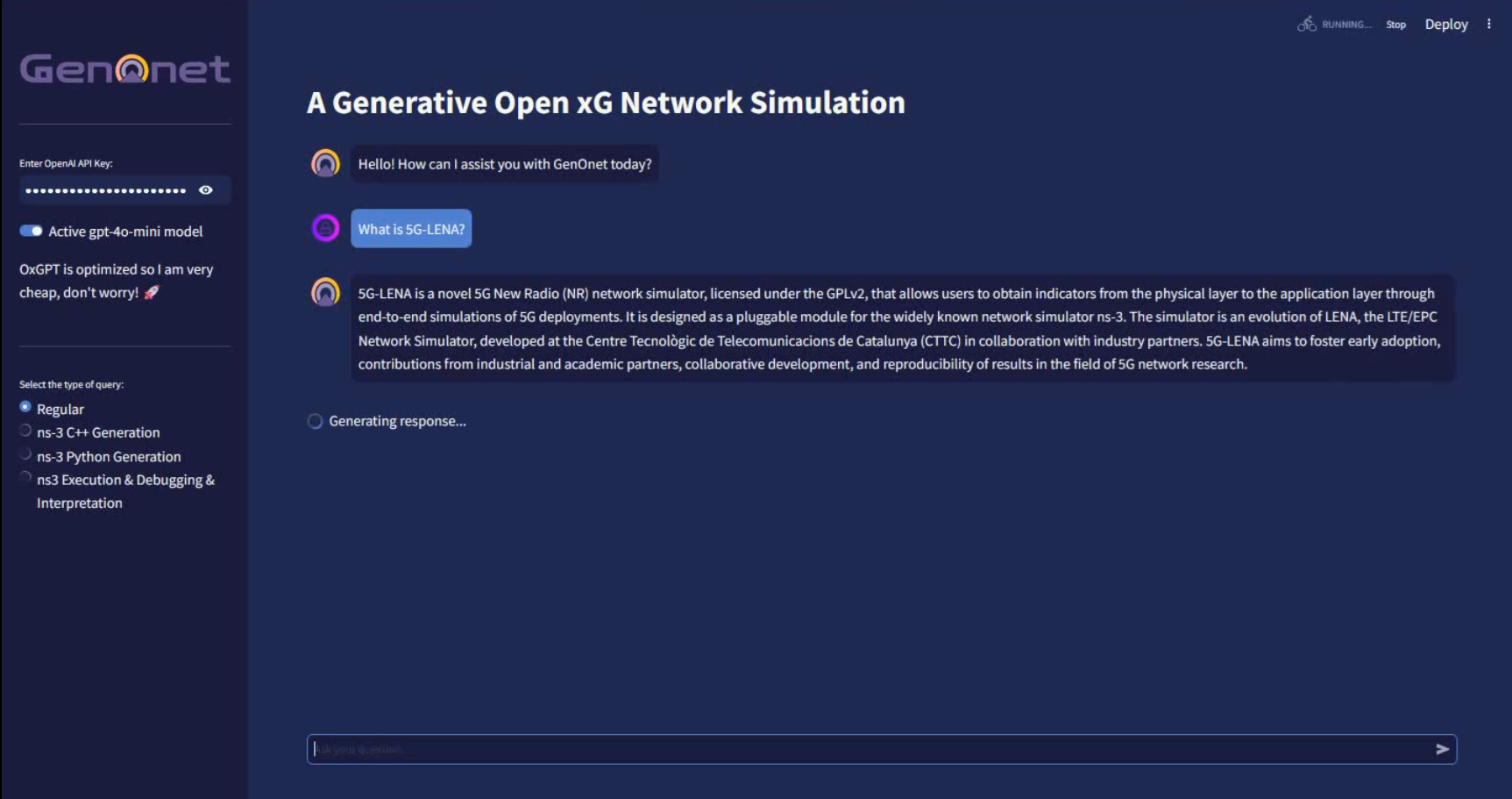}
\caption{The graphical user interface of application.}
\label{fig:gui-llm}

\end{figure}

\begin{figure*}[t]
\centering
\includegraphics[width=1.8\columnwidth, clip,trim={0cm 0cm 0cm 0cm}]{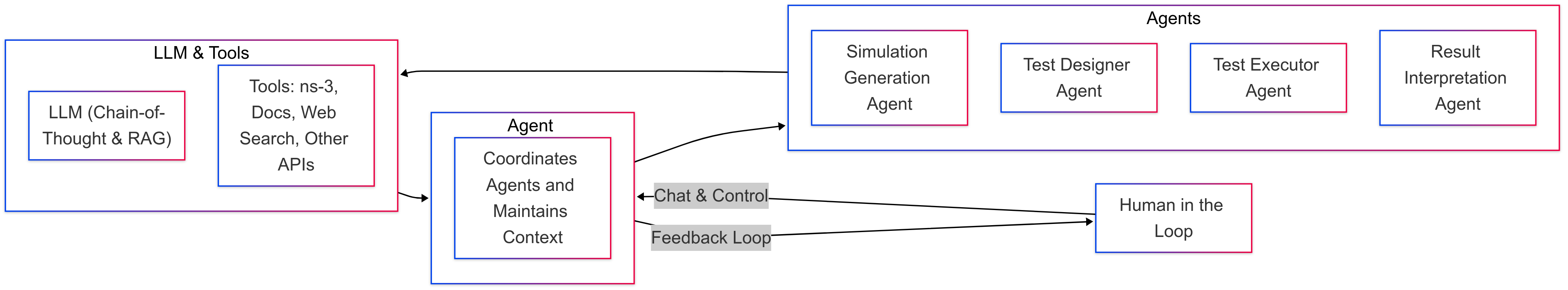}
\caption{Overview of the multi-agent LLM-based ns-3 simulation framework. The streamlined workflow is shown in Figure~\ref{fig:Simulation_workflow}.}
\label{fig:workflow_framework}
\end{figure*}

\section{Methodology}
Figure~\ref{fig:gui-llm} shows the graphical user interface of the proposed framework. This application integrates multiple advanced tools and models within an intuitive and user-friendly environment built using Streamlit. Furthermore, it adopts a modular design approach by utilizing LangChain~\cite{langchainref}and LangGraph~\cite{langgraphref}.

The proposed framework implements a technically sophisticated multi-agent system that leverages language models, retrieval APIs, and custom tool integrations to automate the entire process of network simulation generation, debugging, execution, and analysis in ns-3 environments. Upon receiving simulation requirements in natural language, a central coordinator agent orchestrates the interaction among several specialized agents using LangChain. The Simulation Generation Agent (Agent\#1) utilizes OpenAI's LLMs (accessed via the ChatOpenAI interface) with detailed prompt templates to generate simulation code. In tandem, the Test Designer Agent (Agent\#2) constructs targeted test cases by integrating rule-based logic with retrieval-augmented strategies---utilizing a Pinecone vector store and OpenAIEmbeddings---to verify the simulation's accuracy and reliability. The Test Executor Agent (Agent\#3) subsequently runs the simulation in the ns-3 environment by interfacing with custom execution tools, such as the \texttt{CppSubprocessTool} for C++ code and the \texttt{PythonREPLTool} for Python code. A dynamic toggle in the Streamlit sidebar allows for optimized model selection (e.g., \texttt{gpt-4o-mini}), demonstrating flexible API integration across agents. Finally, the Result Interpretation Agent (Agent \#4) analyzes the simulation outputs by interpreting performance metrics and error logs according to detailed prompt instructions. This actionable feedback is then replied back to the Simulation Generation Agent, which triggers an iterative refinement process. Through this robust integration of API key validations, environment configurations, and multi-tool execution strategies, the framework ensures that the agents collaboratively produce a fully validated and reliable simulation.

Figure~\ref{fig:workflow_framework} shows a comprehensive view of the multi-agent, LLM-driven ns-3 simulation framework. The end user (represented as a human in the loop) provides simulation requirements and feedback to the Agent Orchestration layer, which coordinates specialized agents for simulation generation, test design, execution, and result interpretation. These agents leverage an LLM and external tools (such as ns-3 and documentation) in a cyclical feedback loop to iteratively refine and analyze network simulations.

\subsection{Multi-Agent Workflow for Network Simulation Generation}

\subsubsection{Simulation Generation Agent (Structured Simulation Script Synthesis through Iterative Reasoning)}

The role of the Simulation Generation Agent (Agent\#1) is pivotal in converting high-level simulation specifications into executable ns-3 scripts. This is accomplished by leveraging a structured CoT~\cite{Weicot} reasoning process empowered by LLMs via the ChatOpenAI interface in LangChain. Initially, the agent processes natural language input using advanced Natural language processing (NLP) techniques to extract critical simulation parameters, such as network protocols, mobility models, and bandwidth allocation. Based on these parameters, the agent selects suitable ns-3 libraries and modules (e.g., LTE, Wi-Fi, and 5G-LENA\footnote{\url{https://5g-lena.cttc.es/}}) and identifies optimal simulation models that consider factors like the number of user devices, application demands, environmental conditions, and propagation models (e.g., 3GPP 38.901, COST 231, or ITU-R). In the subsequent script construction phase, the agent automatically generates the simulation script by initializing nodes, configuring network devices, setting up communication channels, and assigning applications. During this process, it instantiates complex ns-3 classes such as \textit{SpectrumWifiPhy} for Wi-Fi frequency modeling and \textit{NrHelper} for mmWave-based networks and leverages integrated tool APIs (e.g., \texttt{CppSubprocessTool} for C++ code execution and \texttt{PythonREPLTool} for Python code debugging) to ensure that each component is rigorously validated. After the initial script has been generated, the Test Designer Agent (Agent\#2) is activated to conduct static analysis and syntax validation. This process helps identify missing dependencies, deprecated API usage, and parameter mismatches. This iterative refinement process ensures that the generated script adheres to ns-3's coding standards and simulation paradigms.

\subsubsection{Test Designer Agent (Rigorous Validation through Automated Test Suite Generation)}

The Test Designer Agent (Agent\#2) validates the generated simulation scripts by automatically constructing a comprehensive suite of test cases. Leveraging LangChain for knowledge retrieval and LLMs for test script generation, this agent creates both primary and edge test cases. Primary test cases verify standard operations such as the successful attachment of user equipments (UEs) to base stations (e.g., eNodeB/gNodeB), the correct data flow between nodes, and adherence to predefined quality of service (QoS) parameters. Edge test cases, on the other hand, guide the simulation to stress conditions such as high mobility speeds, extreme interference, or increased network load to assess network capacity and reliability. Utilizing retrieval-augmented strategies through a Pinecone vector store and OpenAIEmbeddings, the Test Designer Agent dynamically incorporates relevant ns-3 testing paradigms and interfaces with ns-3's testing frameworks (or custom validation code) to ensure comprehensive coverage.

\subsubsection{Test Executor Agent (Dynamic Simulation Execution and Intelligent Feedback Loop)}

The Test Executor Agent (Agent\#3) acts as the operational backbone by executing validated simulation scripts within the ns-3 environment. This agent manages the deployment of simulation scripts across local, virtualized, or cloud-based environments. It captures all dependencies and environment variables to ensure compatibility with the current version of ns-3. During simulation execution, the agent collects diverse output data---including trace files, logs, and performance metrics parsing .pcap files, FlowMonitor outputs, and custom trace sources to extract key performance indicators (KPIs) such as throughput, latency, packet loss, and jitter. Furthermore, advanced NLP techniques and LLM-based analysis are employed to interpret error logs and warnings (e.g., segmentation faults, assertion failures, or unexpected behaviors), enabling the agent to generate detailed, actionable feedback. This feedback is communicated back to the Simulation Generation Agent, establishing an intelligent, iterative feedback loop that drives continuous improvement of the simulation scripts.

\subsubsection{Result Interpretation Agent ( Analyzing Simulation Outcomes and Offering Insights)}

The Result Interpretation Agent (Agent\#4) focuses on post-execution analysis by interpreting the outputs of ns-3 simulations. After a simulation run, the agent processes performance metrics and log data to provide a detailed network behavior analysis. The agent correlates observed performance metrics with underlying network conditions by utilizing sophisticated LLMs with tailored prompt instructions. For example, it may be determined that an increase in end-to-end delay is attributable to a high packet loss rate at the physical layer, potentially caused by wireless channel interference or suboptimal modulation schemes. By delivering clear, actionable insights, the Result Interpretation Agent guides further refinements, ensuring that the simulation meets the desired performance criteria and adheres to best practices in network simulation.
\begin{figure}[h!]
\centering
\includegraphics[width=.9\columnwidth, clip,trim={0cm 0cm 0cm 0cm}]{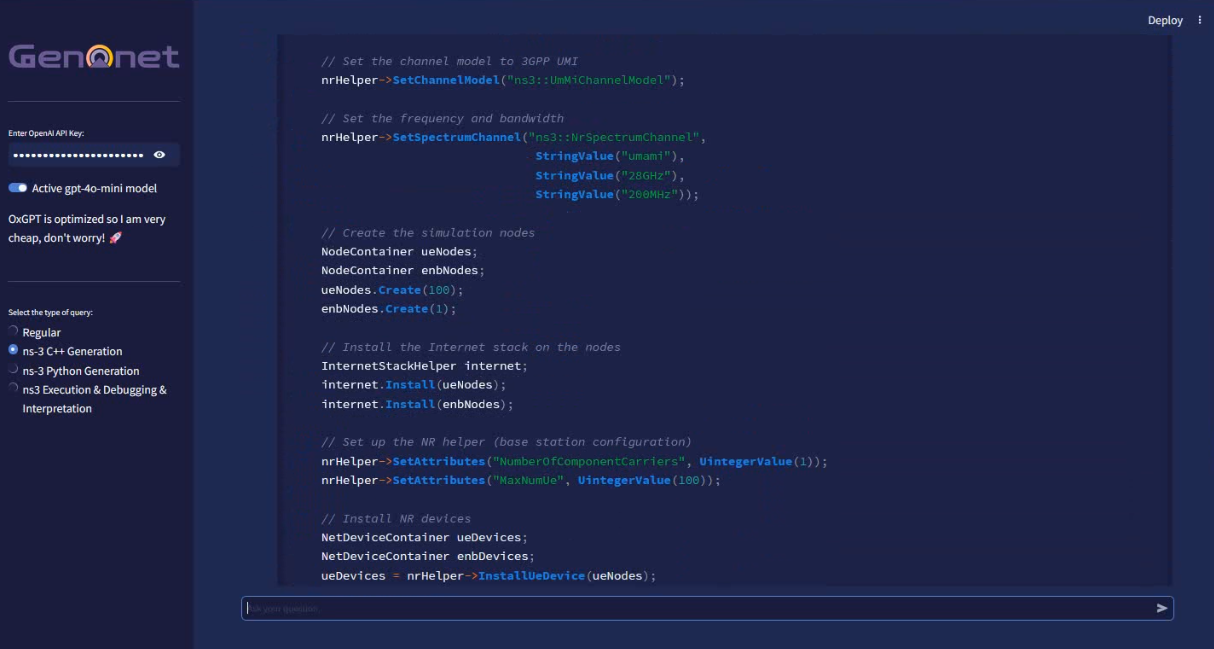}
\caption{The sample ns-3 generated code.}
\label{fig:sample_code}

\end{figure}

\begin{figure*}[t]
\centering
\includegraphics[width=1.8\columnwidth, clip,trim={0cm 0cm 0cm 0cm}]{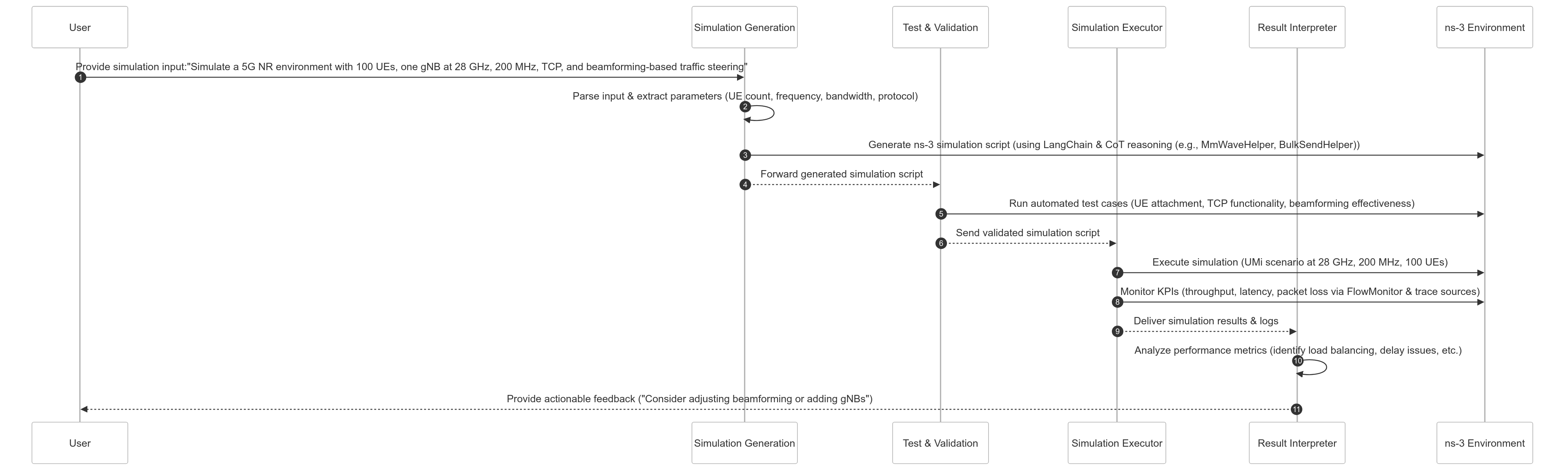}
\caption{Sequence diagram illustrating the streamlined workflow for simulating a 5G NR environment.}
\label{fig:Simulation_workflow}

\end{figure*}

\subsection{Case Study}
This section discusses a specific use case that exemplifies the framework's capabilities in simulating a 5G new radio (NR) environment. The primary objective of this study is to investigate how the proposed framework can facilitate the execution of complex simulation scenarios.

\subsubsection{Simulation Scenario}
The scenario under examination involves simulating a dense urban microcell (UMi) environment characterized by the presence of 100 UEs and one gNB operating at 28 GHz with a bandwidth of 200 MHz. The main focus of this investigation is the implementation of scheduling mechanisms to effectively balance the network load across multiple UEs, thereby ensuring an optimal QoS throughout the network. Figure~\ref{fig:sample_code} demonstrates the setup and configuration of the simulation scenario, including node creation, channel model selection, and Internet stack installation.
The workflow for this use case is outlined in the following steps, which detail the coordinated actions of each agent within the framework.

\subsubsection{Simulation Requirements and Input Parsing (Agent\#1)}  
The user initiates the process by providing a natural language input:  
\textit{"Simulate a 5G New Radio environment with 100 UEs and one gNB at 28 GHz with 200 MHz bandwidth. Implement TCP communication and enable traffic steering using beamforming".} The Agent\#1 receives this input and begins by parsing it using NLP techniques to extract the relevant parameters: frequency, bandwidth, number of UEs, and the desired communication protocol (TCP). By leveraging LangChain, the agent selects appropriate network models from ns-3, such as the 5G-LENA module and 3GPP-based UMi channel models. Using a structured CoT reasoning approach, the agent then synthesizes the necessary ns-3 script. It configures network nodes, assigns UEs, defines the TCP application, and configures the beamforming method. For example, the agent chooses the \texttt{NrHelper} class from ns-3 for mmWave channel modeling and \texttt{BulkSendHelper} for setting up TCP traffic across UEs.

\subsubsection{Test Case Generation and Script Validation (Agent\#2)}  
After the script is generated, Agent\#2 ensures the script's accuracy by creating automated test cases. These tests evaluate the core functionalities of the simulation, such as the correct association of UEs to the gNB, proper functioning of the TCP protocol, and the efficiency of traffic steering using beamforming. Edge cases, such as extreme UE mobility or high interference conditions, are introduced by the agent to test the robustness of the simulation. Scalability tests are also performed, incrementally increasing the number of UEs to assess network performance under heavy traffic loads. The tests automatically interact with ns-3's built-in testing framework to validate these conditions.

\subsubsection{Simulation Execution and Feedback (Agent\#3)}  
With the validation complete, the \textit{Test Executor Agent} (Agent\#3) executes the simulation script within the ns-3 environment, either locally or in a cloud-based setup. The agent ensures that all necessary dependencies and configurations are in place, such as version compatibility with ns-3 modules and the required libraries (e.g., 5G-LENA, TCP/IP). The agent collects performance data during the simulation, such as throughput, latency, and packet loss, by analyzing trace files and logs. It uses FlowMonitor and custom trace sources within ns-3 to gather KPIs. The agent provides detailed feedback if any simulation errors occur, such as a segmentation fault or assertion failure. This information is then passed back to the \textit{Simulation Generation Agent} for iterative refinement of the simulation script, ensuring that the final output is bug-free and meets all performance standards.

\subsubsection{Result Interpretation and Insights (Agent\#4)}  
Once the simulation is successfully executed, the \textit{Result Interpretation Agent} (Agent\#4) analyzes the resulting performance metrics. For this use case, the agent observes that the traffic steering mechanism balances the load efficiently across the UEs, resulting in improved throughput and reduced latency for most UEs. However, the agent also indicates a slight increase in end-to-end delay during periods of heavy traffic, suggesting potential congestion at the gNB. The agent interprets the results and provides actionable feedback, such as:  
\emph{"The increase in delay correlates with a higher packet loss rate during high interference periods, indicating that adjusting the beamforming method or increasing the number of gNBs could alleviate this issue."} Figure~\ref{fig:Simulation_workflow} demonstrates the diagram of interactions between the user, input parsing, test validation, simulation execution, and result interpretation agents, along with the ns-3 environment, highlighting the process from natural language input to actionable feedback.

\section{Experimental Results}
In our experiments, the simulation platform was deployed using a combination of Streamlit for the interface, Pinecone for vector storage, and LangChain and LangGraph integrated with OpenAI models for processing natural language queries. The system was configured to handle multiple queries, including direct simulation commands for Python and C++ environments. Furthermore, the successful initialization of the Pinecone index and the efficient retrieval of relevant information via the question-answering (QA) chain validate the design of our simulation framework. Sample output logs confirmed the correct execution of simulation tasks and indicated system performance. For instance, the output "Index 'ap-v0' created" demonstrates that the vector storage setup was completed without errors. In addition, integrating the ns-3 simulation with custom agents allowed for seamless handling of code generation and execution, further highlighting the system's flexibility.

Table \ref{tab:ingestion_performance} presents the experimental results for the document ingestion process, where the ingestion time increases with the number of documents processed. The data suggests a nearly linear relationship between the number of documents and the processing time, indicating predictable scalability for the ingestion mechanism.

Table \ref{tab:ns3_comparison} compares the execution times for Python and C++ simulation runs. Although both methods ultimately execute the ns-3 simulation, the Python method calls ns-3 using a subprocess (wrapping the ns-3 execution within a Python script), while the C++ method invokes ns-3 directly. The \emph{Base ns-3 Time (s)} represents the inherent execution time of the ns-3 simulation engine when running the simulation code. 

The overhead is defined as the extra time incurred during the invocation process beyond the base simulation time. For the Python invocation, this overhead includes the time required for setting up and managing a subprocess, handling inter-process communications, and any additional interpretation or bridging tasks between Python and the native C++ environment. In contrast, the C++ invocation method directly executes the simulation code, thus introducing significantly less overhead. Table \ref{tab:ns3_comparison} shows that i) The base simulation time provided by ns-3 remains constant (5.0 seconds) and ii) The C++ invocation incurs a smaller overhead (1 second) compared to the Python invocation (3 seconds), leading to slightly lower total execution time. This comparison helps clarify that while the underlying simulation engine is the same, the additional overhead introduced by the invocation method can impact the overall performance.

Table \ref{tab:query_response_time} presents metrics that describe the responsiveness of the system across different types of queries. The \emph{Average Response Time (s)} metric represents the mean time taken for the system to generate a response for each query type, averaged over multiple runs. It indicates the system's efficiency in processing various requests---from regular queries to more complex operations such as code generation and debugging. The~\emph{Standard Deviation (s)} measures the variability or consistency of the response times across multiple trials. A lower standard deviation indicates that the response times are consistently close to the average value, while a higher standard deviation suggests greater fluctuations in the time required to process the queries.

\begin{table}[h]
\centering
\caption{Document Ingestion Performance}
\label{tab:ingestion_performance}
\begin{tabular}{cc}
\hline
\textbf{Number of Documents} & \textbf{Ingestion Time (seconds)} \\
\hline
10   & 1.1 \\
20   & 2.0 \\
30   & 2.9 \\
40   & 3.8 \\
50   & 4.6 \\
60   & 5.4 \\
70   & 6.2 \\
80   & 7.1 \\
90   & 7.9 \\
100  & 8.7 \\
\hline
\end{tabular}

\end{table}

\begin{table*}[ht]
\centering
\caption{Comparison of ns-3 Simulation Execution Performance for Python and C++ Invocation}
\label{tab:ns3_comparison}
\begin{tabular}{lccc}
\hline
\textbf{Invocation Method} & \textbf{Base ns-3 Time (s)} & \textbf{Overhead (s)} & \textbf{Total Time (s)} \\
\hline
Python Invocation (via subprocess) & 5.0 & 3.0 & 8.0 \\
C++ Invocation (direct)             & 5.0 & 1.0 & 6.0 \\
\hline
\end{tabular}

\end{table*}

\begin{table*}[h]
\centering
\caption{Query Response Time Performance}
\label{tab:query_response_time}
\begin{tabular}{lcc}
\hline
\textbf{Query Type} & \textbf{Avg. Response Time (s)} & \textbf{Std. Dev. (s)} \\
\hline
Regular Query & 1.2 & 0.3 \\
ns-3 C++ Generation & 2.7 & 0.4 \\
ns-3 Python Generation & 3.2 & 0.5 \\
ns3 Execution \& Debugging \& Interpretation & 4.5 & 0.6 \\
\hline
\end{tabular}
\end{table*}

\begin{table*}[h]
\centering
\caption{LLM-Driven Code Generation Performance Metrics}
\label{tab:llm_performance}
\begin{tabular}{lccccc}
\hline
\textbf{Simulation Scenario} & \textbf{Avg. Iterations} & \textbf{Syntax Error Rate (\%)} & \textbf{Avg. Response Time (s)} & \textbf{Human Eval Score} &
\textbf{Pass Rate}\\
\hline
Defined Scenario       & 1.8 & 17.0  & 7.3 & 7.5 & 0.72 \\
\hline
\end{tabular}

\end{table*}

The performance metrics summarized in Table~\ref{tab:llm_performance} provide a comprehensive evaluation of the LLM-driven code generation process for the simulation scenario. The \emph{Average Iterations} metric indicates the mean number of iterative refinement cycles required by the LLM to generate syntactically correct and functionally valid ns-3 simulation code. A lower average iteration count implies that the LLM's initial output is close to the desired final form, necessitating minimal corrections. The \emph{Syntax Error Rate (\%)} is calculated as the percentage of syntactical errors present in the generated code relative to the total lines or segments of code produced. A low syntax error rate implies that the LLM is effective at generating code that adheres to ns-3's syntax and coding standards. The \emph{Human Evaluation Score} as a qualitative metric, typically assessed on a scale from 1 to 10, reflects expert judgments regarding the clarity, correctness, and overall quality of the generated code~\cite{humanevalref}. A higher score corresponds to better code quality and usability, as determined by human evaluators. Furthermore, the \emph{Pass@k}~\cite{passatkref} metric is used to quantitatively evaluate code generation models by measuring the probability that at least one of the top-$k$ generated code samples successfully passes all test cases. This metric provides an empirical assessment of a model's ability to produce correct and functional code within a given number of attempts.

The \emph{Average Iterations} metric of 1.8 indicates that, on average, the LLM required fewer than two refinement cycles to produce syntactically correct and functionally accurate ns-3 simulation code, demonstrating that the initial outputs were largely accurate with minimal adjustments needed. The \emph{Syntax Error Rate} of 17.0\% confirms a high degree of syntactic correctness and reduces the need for extensive manual debugging. In addition, an \emph{Average Response Time} (in five runs---$k$ = 5) of 7.3 seconds highlights the computational efficiency of the system, ensuring that iterative refinements are performed rapidly. Finally, \emph{Human Evaluation Score} of 7.5 and \emph{Pass Rate} of 0.72 confirm that the generated code meets high standards.

\section{Open Issues and Future study}
\label{sec:challenges}

\subsection{Automated Bug-Free ns-3 Simulations Using LLMs}
LLMs face challenges when generating ns-3 simulation code due to their hierarchical, C++-based structure and the diversity of protocols (transport, routing, MAC, PHY) maintained by various institutions. Although LLMs may produce syntactically similar code, they often lack a deep understanding of the networking principles needed for correct logic, requiring expert interaction to improve precision.


\subsection{Architectural Enhancements for Protocol-Specific Code Generation}
Current transformer-based LLMs, tailored for natural language and general-purpose coding, struggle with network protocols' highly structured and syntax-driven nature. Incorporating concepts from compiler theory, such as intermediate representations and abstract syntax trees, could better align these models with the requirements of ns-3 code generation.


\subsection{Enhancing Support for Domain-Specific Languages in ns-3 Simulations}
LLMs are typically trained on popular programming languages, leaving domain-specific languages (DSLs) and low-level constructs, which are crucial for ns-3 simulations, less represented. Enhancing DSL support would enable more precise automation of ns-3 code generation.

\subsection{Continuous Learning for Adapting to Evolving Network Standards}
Static LLMs risk becoming outdated with continuous evolution of network standards (e.g., 3GPP, IEEE) and frequent updates to ns-3 modules. Implementing continuous learning mechanisms ensures that generated code remains current and applicable to the latest network technologies.

\section{Conclusion}
The paper presents a multi-agent framework that integrates LLMs with the ns-3 network simulator to automate the creation, validation, execution, and analysis of 5G/6G simulation scenarios. By dividing the process into specialized agents---for generating simulation scripts, designing tests, executing simulations, and interpreting results---the framework transforms natural language inputs into fully validated simulation scripts. Experimental validation in a 5G NR scenario demonstrates its potential to reduce complexity and speed up simulation workflows. However, challenges such as ensuring bug-free code generation, scaling full-stack simulations, and adapting to evolving network standards remain, paving the way for future research.

\end{document}

%% file: acronyms.tex
\newacronym{3gpp}{3GPP}{3rd Generation Partnership Project}
\newacronym{5g}{5G}{Fifth-Generation}
\newacronym{5gc}{5GC}{5G Core}
\newacronym{adc}{ADC}{Analog to Digital Converter}
\newacronym{afbw}{AFBW}{Average Fading Bandwidth}
\newacronym{aimd}{AIMD}{Additive Increase Multiplicative Decrease}
\newacronym{am}{AM}{Acknowledged Mode}
\newacronym{amc}{AMC}{Adaptive Modulation and Coding}
\newacronym{aoa}{AoA}{Angle of Arrival}
\newacronym{aod}{AoD}{Angle of Departure}
\newacronym{ap}{AP}{Access Point}
\newacronym{app}{APP}{Application Layer}
\newacronym{aqm}{AQM}{Active Queue Management}
\newacronym{awgn}{AGWN}{Additive White Gaussian Noise}
\newacronym{balia}{BALIA}{Balanced Link Adaptation}
\newacronym{bdp}{BDP}{Bandwidth-Delay Product}
\newacronym{ber}{BER}{Bit Error Rate}
\newacronym{bler}{BLER}{Block Error Rate}
\newacronym{bf}{BF}{Beamforming}
\newacronym{cad}{CAD}{Computer-Aided Design}
\newacronym{cbr}{CBR}{Constant Bit Rate}
\newacronym{qos}{QOS}{Quality of Service}
\newacronym{cc}{CC}{Congestion Control}
\newacronym{cdf}{$CDF$}{Cumulative Distribution Function}
\newacronym{ci}{CI}{Confidence Interval}
\newacronym{oran}{O-RAN}{Open Radio Access Network}
\newacronym{cir}{CIR}{Channel Impulse Response}
\newacronym{cn}{CN}{Core Network}
\newacronym{cp}{CP}{Control Plane}
\newacronym{cqi}{CQI}{Channel Quality Information}
\newacronym{crs}{CRS}{Cell Reference Signal}
\newacronym{csirs}{CSI-RS}{Channel State Information - Reference Signal}
\newacronym{dc}{DC}{Dual Connectivity}
\newacronym{dce}{DCE}{Direct Code Execution}
\newacronym{dci}{DCI}{Downlink Control Information}
\newacronym{llm}{LLM}{Large Language Model}
\newacronym{dl}{DL}{Downlink}
\newacronym{dmr}{DMR}{Deadline Miss Ratio}
\newacronym{dmrs}{DMRS}{DeModulation Reference Signal}
\newacronym{dray}{D-Ray}{Deterministic Ray}
\newacronym{e2e}{E2E}{End-to-End}
\newacronym{ecn}{ECN}{Explicit Congestion Notification}
\newacronym{ecdf}{ECDF}{Empirical Cumulative Distribution Function}
\newacronym{edf}{EDF}{Earliest Deadline First}
\newacronym{em}{EM}{electromagnetic}
\newacronym{enb}{eNB}{evolved Node Base}
\newacronym{endc}{EN-DC}{E-UTRAN-\gls{nr} \gls{dc}}
\newacronym{epc}{EPC}{Evolved Packet Core}
\newacronym{es}{ES}{Edge Server}
\newacronym{fdd}{FDD}{Frequency Division Duplexing}
\newacronym{fdma}{FDMA}{Frequency Division Multiple Access}
\newacronym{fray}{F-Ray}{Flashing Ray}
\newacronym{fs}{FS}{Fast Switching}
\newacronym{ftp}{FTP}{File Transfer Protocol}
\newacronym{gmm}{GMM}{Gaussian Mixture Model}
\newacronym{gnb}{gNB}{Next Generation Node Base}
\newacronym{harq}{HARQ}{Hybrid Automatic Repeat reQuest}
\newacronym{hetnet}{HetNet}{Heterogeneous Network}
\newacronym{hh}{HH}{Hard Handover}
\newacronym{hol}{HOL}{Head-of-Line}
\newacronym{hqf}{HQF}{Highest-quality-first}
\newacronym{ia}{IA}{Initial Access}
\newacronym{iab}{IAB}{Integrated Access and Backhaul}
\newacronym{ieee}{IEEE}{Institute of Electrical and Electronics Engineers}
\newacronym{imt}{IMT}{International Mobile Telecommunication}
\newacronym{inr}{INR}{Interference to Noise Ratio}
\newacronym{iot}{IoT}{Internet of Things}
\newacronym{ked}{KED}{Knife-Edge Diffraction}
\newacronym{kpi}{KPI}{Key Performance Indicator}
\newacronym{oxgpt}{GenOnet}{Generative Open xG Network Simulation}
\newacronym{ks}{KS}{Kolmogorov–Smirnov}
\newacronym{lcf}{LCF}{Level Crossing Frequency}
\newacronym{lcr}{LCR}{Level Crossing Rate}
\newacronym{los}{LoS}{Line-of-Sight}
\newacronym{lte}{LTE}{Long Term Evolution}
\newacronym{m2m}{M2M}{Machine to Machine}
\newacronym{mac}{MAC}{Medium Access Control}
\newacronym{mc}{MC}{Multi-Connectivity}
\newacronym{mcs}{MCS}{Modulation and Coding Scheme}
\newacronym{mec}{MEC}{Mobile Edge Cloud}
\newacronym{mi}{MI}{Mutual Information}
\newacronym{mib}{MIB}{Master Information Block}
\newacronym{mimo}{MIMO}{Multiple Input, Multiple Output}
\newacronym{mlr}{MLR}{Maximum-local-rate}
\newacronym[plural=\gls{mme}s,firstplural=Mobility Management Entities (MMEs)]{mme}{MME}{Mobility Management Entity}
\newacronym{mmwave}{mmWave}{millimeter wave}
\newacronym{moi}{MoI}{Method of Images}
\newacronym{mpc}{MPC}{Multi Path Component}
\newacronym{mptcp}{MPTCP}{Multipath TCP}
\newacronym{mr}{MR}{Maximum Rate}
\newacronym{mrdc}{MR-DC}{Multi \gls{rat} \gls{dc}}
\newacronym{mss}{MSS}{Maximum Segment Size}
\newacronym{mtd}{MTD}{Machine-Type Device}
\newacronym{mtu}{MTU}{Maximum Transmission Unit}
\newacronym{nfv}{NFV}{Network Function Virtualization}
\newacronym{nist}{NIST}{National Institute of Standards and Technology}
\newacronym{nlos}{NLoS}{Non-Line-of-Sight}
\newacronym{nr}{NR}{New Radio}
\newacronym{nrmse}{NRMSE}{Normalized Root Mean Square Error}
\newacronym{ns3}{ns-3}{network simulator 3}
\newacronym{nsa}{NSA}{Non Stand Alone}
\newacronym{o2i}{O2I}{Outdoor-to-Indoor}
\newacronym{ofdm}{OFDM}{Orthogonal Frequency Division Multiplexing}
\newacronym{pa}{PA}{Position-aware}
\newacronym{pbch}{PBCH}{Physical Broadcast Channel}
\newacronym{pdcch}{PDCCH}{Physical Downlonk Control Channel}
\newacronym{pdcp}{PDCP}{Packet Data Convergence Protocol}
\newacronym{pdsch}{PDSCH}{Physical Downlink Shared Channel}
\newacronym{pdu}{PDU}{Packet Data Unit}
\newacronym{per}{PER}{Packet Error Rate}
\newacronym{pf}{PF}{Proportional Fair}
\newacronym{pgw}{PGW}{Packet Gateway}
\newacronym{phy}{PHY}{Physical}
\newacronym{pl}{PL}{Path Loss}
\newacronym{ppp}{PPP}{Poisson Point Process}
\newacronym{prb}{PRB}{Physical Resource Block}
\newacronym{pss}{PSS}{Primary Synchronization Signal}
\newacronym{pucch}{PUCCH}{Physical Uplink Control Channel}
\newacronym{pusch}{PUSCH}{Physical Uplink Shared Channel}
\newacronym{qd}{QD}{Quasi Deterministic}
\newacronym{rach}{RACH}{Random Access Channel}
\newacronym{ran}{RAN}{Radio Access Network}
\newacronym[firstplural=Radio Access Technologies (RATs)]{rat}{RAT}{Radio Access Technology}
\newacronym{red}{RED}{Random Early Detection}
\newacronym{rf}{RF}{Radio Frequency}
\newacronym{fr}{FR}{Frequency Range}
\newacronym{rlc}{RLC}{Radio Link Control}
\newacronym{rlf}{RLF}{Radio Link Failure}
\newacronym{rr}{RR}{Round Robin}
\newacronym{rray}{R-Ray}{Random Ray}
\newacronym{rrc}{RRC}{Radio Resource Control}
\newacronym{rrm}{RRM}{Radio Resource Management}
\newacronym{rs}{RS}{Remote Server}
\newacronym{rsrp}{RSRP}{Reference Signal Received Power}
\newacronym{rsrq}{RSRQ}{Reference Signal Received Quality}
\newacronym{rss}{RSS}{Received Signal Strength}
\newacronym{rssi}{RSSI}{Received Signal Strength Indicator}
\newacronym{rt}{RT}{Ray Tracer}
\newacronym{rtt}{RTT}{Round Trip Time}
\newacronym{rw}{RW}{Receive Window}
\newacronym{rx}{RX}{Receiver}
\newacronym{sa}{SA}{standalone}
\newacronym{sack}{SACK}{Selective Acknowledgment}
\newacronym{sap}{SAP}{Service Access Point}
\newacronym{sch}{SCH}{Secondary Cell Handover}
\newacronym{scm}{SCM}{Stochastic Channel Model}
\newacronym{scoot}{SCOOT}{Split Cycle Offset Optimization Technique}
\newacronym{sdma}{SDMA}{Spatial Division Multiple Access}
\newacronym{sf}{SF}{Shadow Fading}
\newacronym{si}{SI}{Study Item}
\newacronym{sib}{SIB}{Secondary Information Block}
\newacronym{sinr}{SINR}{Signal-to-Interference-plus-Noise Ratio}
\newacronym{sir}{SIR}{Signal-to-Interference Ratio}
\newacronym{sm}{SM}{Saturation Mode}
\newacronym{snr}{SNR}{Signal-to-Noise Ratio}
\newacronym{son}{SON}{Self-Organizing Network}
\newacronym{srs}{SRS}{Sounding Reference Signal}
\newacronym{ss}{SS}{Synchronization Signal}
\newacronym{sss}{SSS}{Secondary Synchronization Signal}
\newacronym{sta}{STA}{Station}
\newacronym{svd}{SVD}{Singular Value Decomposition}
\newacronym{tb}{TB}{Transport Block}
\newacronym{tcp}{TCP}{Transmission Control Protocol}
\newacronym{udp}{UDP}{User Datagram Protocol}
\newacronym{tdd}{TDD}{Time Division Duplexing}
\newacronym{tdma}{TDMA}{Time Division Multiple Access}
\newacronym{tfl}{TfL}{Transport for London}
\newacronym{tgad}{TGad}{Task Group ad}
\newacronym{tgay}{TGay}{Task Group ay}
\newacronym{tm}{TM}{Transparent Mode}
\newacronym{trp}{TRP}{Transmitter Receiver Pair}
\newacronym{tti}{TTI}{Transmission Time Interval}
\newacronym{ttt}{TTT}{Time-to-Trigger}
\newacronym{tx}{TX}{Transmitter}
\newacronym{ue}{UEs}{User Equipments}
\newacronym{ul}{UL}{Uplink}
\newacronym{um}{UM}{Unacknowledged Mode}
\newacronym{uma}{UMa}{Urban Macro}
\newacronym{uml}{UML}{Unified Modeling Language}
\newacronym{utc}{UTC}{Urban Traffic Control}
\newacronym{vm}{VM}{Virtual Machine}
\newacronym{wbf}{WBF}{Wired Bias Function}
\newacronym{wf}{WF}{Wired-first}
\newacronym{wifi}{Wi-Fi}{Wireless Fidelity}
\newacronym{wigig}{WiGig}{Wireless Gigabit}
\newacronym{wlan}{WLAN}{Wireless Local Area Network}
\newacronym{xpr}{XPR}{Cross Polarization Ratio}
\newacronym{thz}{THz}{terahertz}
\newacronym{ghz}{GHz}{Gigahertz}
\newacronym{ai}{AI}{Artificial Intelligence}
\newacronym{dnn}{DNN}{Deep Neural Network}
\newacronym{6g}{6G}{Sixth-Generation}
\newacronym{toa}{ToA}{Time of Arrival}
\newacronym{fsc}{FS}{Fully Stochastic}
\newacronym{hbc}{HB}{Hybrid}
\newacronym{hpbw}{HPBW}{Half Power Beamwidth}
\newacronym{tc}{TC}{Time Cluster}
\newacronym{sl}{SL}{Spatial Lobe}
\newacronym{gan}{GAN}{Generative Adversial Network}
\newacronym{cgan}{cGAN}{Conditional Generative Adversial Network}
\newacronym{relu}{$ReLU$}{Rectified Linear Unit}
\newacronym{mae}{$MAE$}{Mean Absolute Error}
\newacronym{SSCH-THz}{SSCH-THz}{Simplified Stochastic Channel Model in THz}
\newacronym{vr}{VR}{Virtual Reality}
\newacronym{edf1}{EDF}{Empirical Distribution Function}
\newacronym{sse}{SSE}{Sum Square Error}
\newacronym{soa}{SOA}{State Of Art}
\newacronym{ric}{RIC}{RAN Intelligent Controller}
\newacronym{lm}{LM}{Location Manager}
\newacronym{umi}{UMi}{urban microcells}
\newacronym{xapp}{xApp}{eXtended application}
\newacronym{ntn}{NTN}{non-terrestrial network}
\newacronym{nlp}{NLP}{Natural Language Processing}